\pdfoutput=1
\documentclass[?4,10pt,onecolumn]{article}
\usepackage{color}
\usepackage{graphicx}
\usepackage{amsmath,amssymb}
\usepackage{multirow}
\usepackage{url}
\usepackage{physics}

\newcommand{\rot}{{\rm rot}}

\begin{document}

\title{Three-dimensional calculations of the inductive coupling between radio-frequency waves and plasma in the drivers of the SPIDER device}
\author{D. L\'opez-Bruna$^1$, M. Recchia$^2$, P. Jain$^2$,  I. Predebon$^{2,3}$\\ S. Denizeau$^2$, A. La Rosa$^2$}
\date{\footnotesize 
\begin{flushleft}$^1$ Laboratorio Nacional de Fusi\'on - CIEMAT, Madrid, Spain\\
$^2$ Consorzio RFX (CNR, ENEA, INFN, Universit\`a di Padova, Acciaierie Venete SpA), Padova, Italy\\
$^3$ Istituto per la Scienza e Tecnologia dei Plasmi, CNR, Padova, Italy
\end{flushleft}}

%%%%%%%%%%%%%%%%%%%% NOTA %%%%%%%%%%%%%%
%
%      Busca "XXXXX" para ver d—nde hay que acabar o repasar el texto
%
%%%%%%%%%%%%%%%%%%%% NOTA %%%%%%%%%%%%%%

\maketitle

\abstract
{\small 
This work documents the initial 3D calculations to simulate the coupling between radio-frequency (RF) waves and plasma in discharges of the SPIDER device. Axisymmetric 3D calculations in the plasma domain alone compare well against equivalent 2D cases. A model of SPIDER driver, the cylindrical chamber where the plasma is heated by the RF drive, is then defined including the metallic parts of the Faraday shield, insulator and vacuum layer up to the RF winding (not included in the calculation domain). Estimates of the power share in the different parts are obtained using experimental conditions and plasma data. The results are sensitive to the particular geometry of the driver and the temperature of the Faraday shield, but generally agree with the experimental knowledge. The ratio between total delivered power and plasma absorbed power is found, depending on the plasma parameters, in the range 30--45\%.}

\section{Introduction}

Neutral Beam Injection (NBI) heating has been adopted as the main procedure to heat up magnetically confined fusion plasmas, namely the ITER device. The technique consists in creating a small plasma from which ions are extracted and accelerated up to the desired energy, and then neutralized to make a beam of neutral particles able to penetrate the magnetic field and transfer its energy to the fusion plasma. The construction and commissioning of NBIs for ITER is an acknowledged technical challenge. Indeed, the Neutral Beam Test Facility (NBTF) at Padua was started with the purpose of delivering a tested final design of the ITER NBI system \cite{esch2015physics-design-,hemsworth2017overview-of-the}. Two main experiments are being developed in the NBTF: the one-to-one scale NBI source prototype for ITER, MITICA \cite{agostinetti2015detailed-design}; and the NBI source, also at full scale, including the plasma production and sustainment and the beam formation, SPIDER \cite{Serianni2019SPIDER-in-the-r}. The source plasma is produced in a set of eight cylindrical cavities surrounded by high-power induction coils, called ``drivers'', where the induced fields transmit energy mainly to the electrons in order to heat up and sustain the plasma. Hence the name Inductively Coupled Plasma (ICP). SPIDER experiments, which started in 2018 \cite{Chitarin2018Start-of-SPIDER}, aim at meeting the demanding requirements of negative ion current and beam divergence during long ($\sim$ 1 h) pulses in the ITER NBI sources. During SPIDER operation it has become clear the need for proper modelling of the plasma properties in the NBI source: adequate plasma parameters must be achieved in order to guarantee the quality of the extracted beam in terms of achievable accelerated negative ion currents, homogeneity of the beams in the large extraction grid, etc. This calls for plasma transport estimates in the drivers and expansion region towards the plasma and acceleration grids. In turn, heat and particle sources are essential ingredients in any plasma transport calculation. This work concentrates on the heat source in the drivers of the SPIDER device. In particular, we are interested in modelling the coupling of RF waves produced by the induction coils with the plasma and surrounding metallic parts in the driver region.

A basic figure of merit of an ICP source is the Power Transfer Efficiency (PTE), or percentage of the net power delivered to the driver that couples to the plasma.  Previous works \cite{Jain2018Improved-Method,Recchia2021studies-on-powe} have been devoted to estimate the PTE of the drivers in the SPIDER device based on the near cylindrical symmetry of the problem. Mutual inductances for a large set of circular loops of currents (coils, other driver metallic parts and the plasma itself) are used to reduce the induction problem to a set of linear equations from which the currents, and hence the ohmic power dissipated in the driver, can be obtained. In these works, basic 0D transport considerations allow for a first degree of self-consistency, i.e., the power as external input gives rise to different couplings to the plasma depending on the characteristics of the latter, which in turn depend on the deposited power. There is ongoing research devoted to study the transport properties of SPIDER plasmas, both from a 2D-fluid perspective \cite{zagorski20222-d-fluid-model,zagorski20232d-simulations} and from a 3D description using ``particle in cell'' techniques for specific purposes \cite{candeloro2022electron-scrapi}. These transport codes can greatly improve the self-consistent treatment of the PTE problem as long as the ICP part is properly solved. Since the technical details of the drivers (like different materials with not necessarily cylindrical symmetry, and stray magnetic fields) and the electrical properties of the plasma make this a complex study, different approaches to the ICP problem and an appropriate cross-check are advisable. This motivated starting also an electromagnetic treatment based on solving directly Maxwell's equations on a medium given by the experimental electron density and temperature.   A 2D code was developed \cite{lopezbruna20212Delectromagnet} to obtain first estimates of the induced electric and magnetic fields inside the drivers assuming perfect axi-symmetry, i.e: the induced electric field and  plasma current density only have azimuthal component around the driver axis. These calculations confirmed the need of including a strong reduction of the electrical conductivity in presence of an RF magnetic field \cite{Recchia2021studies-on-powe,jain2022investigation-o}. 2D-fluid transport calculations that include the same models for ICP in a self-consistent manner \cite{zagorski20222-d-fluid-model,zagorski20232d-simulations} also demand some kind of effective plasma conductivity much reduced where the magnitude of the RF magnetic field is larger. All these calculations assume the approximation of perfect axi-symmetry, a condition that is broken,  due to the particle drifts,  by the plasma itself when there is a transverse static magnetic field; and, inevitably, by the Faraday shield, a metallic structure that isolates the plasma and minimizes undesired capacitive effects.

\begin{figure}[htbp] %  figure placement: here, top, bottom, or page
   \centering
   \includegraphics[width=0.5\textwidth]{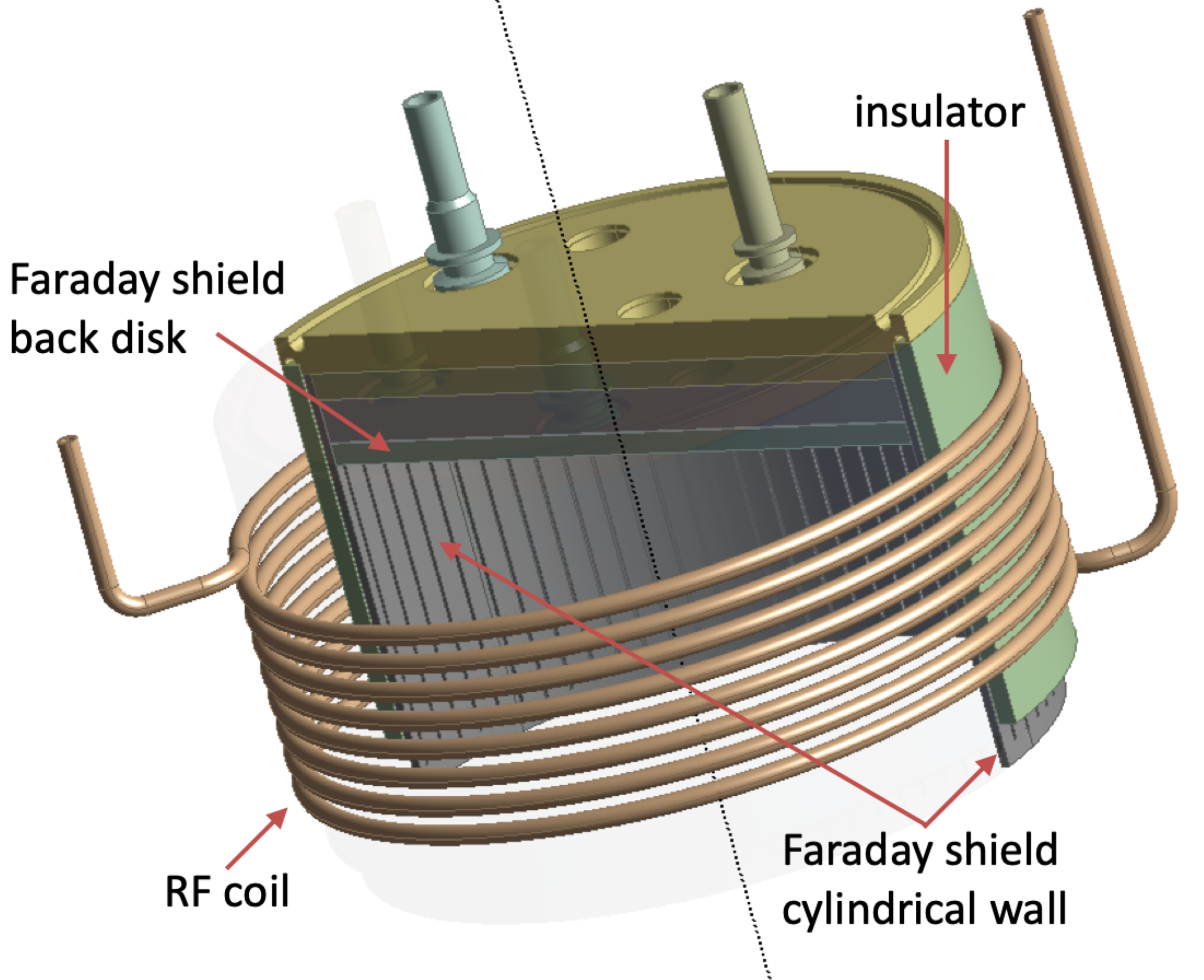} 
   \caption{CAD drawing of a SPIDER driver. A dotted line marks the axes of the cylindrical structure from which a radial coordinate can be defined. The bottom part opens towards the plasma expansion region.}
   \label{fig:real driver}
\end{figure}

Figure \ref{fig:real driver} shows a computer drawing of any of the eight SPIDER drivers. The image shows all the main elements relevant for  ICP calculations: the insulator, the back disk and the Faraday shield enclosing the plasma volume. A cylindrical electromagnetic shield (not shown) covers the RF coil and isolates the ensemble from neighboring drivers and other components of the source. The 3D calculations to be presented include neither the RF coil nor the electromagnetic shield, which are basically axisymmetric and amenable to 2D calculations from which the boundary conditions for the 3D computations are taken. This will be explained later on.

A 3D calculation of the ICP should serve to evaluate the importance of intrinsically 3D effects, like the presence of an external magnetic field or non cylindrically-symmetric plasma inhomogeneities. In the end, the purpose is to develop a reliable tool to study the inductive coupling, or interplay between the plasma and the electromagnetic RF fields, considering all the relevant elements of the plasma source. As a necessary step before attempting the modelling of 3D effects, we have envisaged a comparison with previous electromagnetic 2D calculations. Therefore, this work has two initial objectives: providing (i) documentation about the 3D calculation method for the ICP in the drivers of SPIDER and (ii) a comparison with axisymmetric calculations. As an outcome of these calculations, we obtain (iii) the share of power in the main different parts of the driver and the incidence of the copper temperature in the PTE; and (iv) a scaling of the losses in the Faraday shield with the current in the RF coils, i.e., the effective resistance associated to the Faraday shield.

With the objectives above in mind, we have organized the rest of this work as follows. In Section \ref{sec:ecuaciones} we show the equations that we are solving (\ref{subsec:ecuacion}), then give some indications about the numerical method (\ref{subsec:formulacion variacional}), the boundary conditions (\ref{subsec:condiciones de contorno}) and the numerical integration (\ref{subsec:integracion}). % and, finally, the restriction to 2D (\ref{subsec:2D version}).
 In Section \ref{sec:cazuela} we describe the model we have adopted for the plasma surrounding elements up to the calculation limits, just before the RF coil in radial coordinate. These two sections refer to the objective (i) above. Section \ref{sec:resultados} contains the results of the calculations.  In  (\ref{subsec:compara 2D}) we make a comparison with previous 2D models, where we show that they have been indeed transported to 3D (objective (ii)). A scan to show the sensitivity of the power share depending on the number of slits of the Faraday shield and their spacing is presented in (\ref{subsec:rendijas}); and the power share in the different parts depending on the temperature associated to the Faraday-shield lateral wall comes in  (\ref{subsec:temperatura cobre}), thus dedicated to the objectives (iii) and (iv). The paper finishes with a brief conclusion.

\section{\label{sec:ecuaciones}Equations and numerical setup}

For completeness, this section is dedicated to give a quick overview of the equations to be solved and the methods chosen to find numerical solutions.

\subsection{\label{subsec:ecuacion}Equation}

Let us recall briefly the equations that lead to the one to be solved. The problem involves (i) the magnetic field intensity, whose sources are material and displacement current densities, $\curl \mathbf{H} = \mathbf{J} + \partial_t \mathbf{D}$; and (ii) the induction process, which implies a vortex source for the induced electric field due to time-varying magnetic fields, $\curl \mathbf{E} = -\partial_t \mathbf{B}$. All these fields are taken in the laboratory frame. The materials are characterized macroscopically by the dielectric properties $\epsilon$, the conductivity $\sigma$ and the permeability $\mu$; that is, $\mathbf{D} = \epsilon \mathbf{E}$, $\mathbf{J} = \sigma \mathbf{E} + \mathbf{J}_\mathrm{b}$, and $\mathbf{B} = \mu \mathbf{H}$,
which might present complicated dependencies like being field-dependent or non local. For simplicity, in what follows we consider $\sigma$, $\epsilon$ and $\mu$ scalars. $\mathbf{J}_\mathrm{b}$ is a driving current density field, like the one in the RF winding. We thus obtain a typical double-curl problem involving material,  immaterial and imposed current densities,
\begin{equation}
 \curl \curl \mathbf{A} =  \mu [\sigma \mathbf{E} + \partial_t (\epsilon \mathbf{E}) +  \mathbf{J}_\mathrm{b}].
\label{ec:olla2_0}
\end{equation}
Since in our case  $\mathbf{E}$ is an induced electric field (charge accumulation effects are not considered), we have $\partial_t \mathbf{B} = -\curl \mathbf{E} = \curl \partial_t \mathbf{A}$; so the substitution $\mathbf{E} = -\partial_t \mathbf{A}$ makes the previuos equation become
\begin{equation}
\curl \curl \mathbf{A} + \mu \sigma \partial_t \mathbf{A}  + \mu\epsilon \partial_{tt} \mathbf{A} = \mu \mathbf{J}_\mathrm{b}.
\label{ec:olla2_1}
\end{equation}
We simplify the problem by assuming the stationary case associated to a single frequency where the current density and the vector potential are the real parts of harmonic solutions in the complex plane, respectively $\Re \{\mathbf{J}_\mathrm{b} (\mathbf{x}) e^{\imath \omega t}\}$ and $\Re \{ \mathbf{A}(\mathbf{x})  e^{\imath \omega t} \}$. Therefore we seek the spatial distribution of the amplitudes, $\curl \curl \mathbf{A} (\mathbf{x}) + (\imath  \omega \mu \sigma - \omega^2 \mu \epsilon ) \mathbf{A} (\mathbf{x}) =  \mu \mathbf{J}_\mathrm{b}$. Defining the complex coefficient
\begin{equation}
k (\mathbf{x}) \equiv \imath \mu \omega \sigma (\mathbf{x}) - \omega^2 \mu \epsilon (\mathbf{x}),
\label{ec:k}
\end{equation}
we rewrite equation \ref{ec:olla2_1} in the form that is to be solved numerically,
\begin{equation}
\curl \curl \mathbf{A}  + k \mathbf{A} =  \mathbf{f}.
\label{ec:olla2_6}
\end{equation}
Here  $\mathbf{f} \equiv  \mu \mathbf{J}_\mathrm{b}$, a term that is null in the following 3D calculations because we always solve the equation in a domain that excludes the RF winding. Otherwise, care has to be taken to make it either solenoidal or, in case it is a current that goes through the calculation domain, a scalar potential $\varphi$ should be defined and a term $(\imath \omega \epsilon + \sigma) \grad \varphi$ should be included in the left hand side of Eq.~\ref{ec:olla2_1}, see e.g.~\cite{Hiptmair2008A-Robust-Maxwel}. Exceptionally, $\mathbf{J}_\mathrm{b}$ is included in 2D calculations (Sec.~\ref{subsec:2D version}) precisely to obtain appropriate boundary conditions for the 3D calculations.
 
In some cases the displacement current, $\partial_t (\epsilon \mathbf{E})$ in equation \ref{ec:olla2_0}, can be neglected in comparison with the $\sigma \vb{E}$ term. Since we want to also consider regions where the conductivity is negligible (dielectric, vacuum), we retain the displacement currents in the problem vector equation, which gives rise to the $\omega^2$ term in the coefficient Eq.~\ref{ec:k}. We should keep in mind that the plasma electrical conductivity, especially in presence of static magnetic fields, might be better described by a tensor rather than a scalar; and, furthermore, that a non-linear response of the plasma to the induced electromagnetic fields may not be suitable for a one-harmonic description of the problem like equations \ref{ec:k}--\ref{ec:olla2_6}. As mentioned, we leave these considerations aside in what follows.

\subsection{\label{subsec:formulacion variacional}Variational formulation}

\subsubsection{\label{subsubsec:3D}3D calculations}

Equation \ref{ec:olla2_6} is to be solved in a 3D domain where, in some cases, different material parts will be present with possibly complicated geometries. For this reason we have decided to solve the problem using the Finite Element Method (FEM), which requires a variational formulation. To give a rough idea for those readers less familiar with the topic, let us think of Eq.~\ref{ec:olla2_6} as an example of equation where the differential operator $\mathcal{D} \equiv \rot~\rot + k$ 
acts on a vector function $u$. Here $\rot$ stands for the curl of a vector field and $k$ is a scalar field. If there exists a solution $u$ such that $\mathcal{D} u = f$ for a prescribed vector function $f$, then we expect that the inner product by another arbitrary vector function $v$ satisfies $\langle \mathcal{D} u, v \rangle = \langle f, v \rangle$. Such inner product in the infinite-dimensional space of 3D functions is expressed in terms of volume integrals, generally referred to as functionals (each space function is associated to a scalar). Of course, the interested reader should look for one of the many treaties on the topic for rigorous definitions and explanations. Here, let us just say that the reverse problem is not trivial:  can we guarantee that $u$ is a solution of our differential equation if $\langle \mathcal{D} u, v \rangle = \langle f, v \rangle$ for some \emph{given} family of functions $v$? Mathematics proves that the answer is affirmative for properly prescribed function families and inner products. The FEM is based on solving the system of equations that comes out of $\langle \mathcal{D} u, v \rangle = \langle f, v \rangle$ for suited function families of finite dimension (for instance order-two polynomials) both for $u$ and $v$, defined piecewise in small portions (finite elements) of the domain.

 The particular solution to an equation like \ref{ec:olla2_6} depends on the boundary conditions. These are imposed in the variational problem owing to a generalization of the integration by parts. In order to, once more, give a rapid intuition of the technique, let us just recall that integral theorems, like the divergence theorem in 3D, relate volume integrals with surface integrals where the boundary conditions prevail. This provides a formulation where the boundary conditions are naturally imposed. 
This said, we show how Eq.~\ref{ec:olla2_6} can be transformed in order to include the boundary conditions in our particular variational formulation of the problem. Our inner product is the Euclidean one and we seek a vector-potential function $\vb{A}$ such that, in the 3D domain of interest, $\Omega \subset \mathbb{R}^3$, 
\begin{equation}
\int_\Omega (\curl \curl \mathbf{A} + k \mathbf{A} )\vdot \mathbf{v} \dd^3 \mathbf{x} \equiv \int_\Omega \mathcal{D}  \mathbf{A} \vdot \mathbf{v} \dd^3 \mathbf{x} = \int_\Omega \mathbf{f} \vdot \mathbf{v} \dd^3 \mathbf{x}
\label{ec:olla4_1}
\end{equation}
for, in principle, any vector function $\mathbf{v}(\mathbf{x})$ belonging to the finite dimensional dual space. Recalling the identity $
\mathbf{v}\vdot \curl \mathbf{a} = \curl \mathbf{v} \vdot \mathbf{a} - \div (\mathbf{v} \cross \mathbf{a})
$ we may consider $\mathbf{a} = \curl \mathbf{A}$ and write
\begin{equation}
 \curl \curl \mathbf{A}  \vdot \mathbf{v} = \curl \mathbf{v} \vdot \curl \mathbf{A} - \div (\mathbf{v} \cross \curl \mathbf{A} ).
\label{ec:olla5_3}
\end{equation}
Hence, our problem is equivalent to the following one:
\begin{equation}
\int_\Omega \curl  \mathbf{A} \vdot \curl \mathbf{v} ~\dd^3 \mathbf{x} + \int_\Omega  k \mathbf{A} \vdot \mathbf{v} ~\dd^3 \mathbf{x} =  \int_\Omega \mathbf{f} \vdot \mathbf{v} ~\dd^3 \mathbf{x} + \int_{\partial \Omega} \mathbf{v} \vdot \curl  \mathbf{A}  \cross \hat{\mathbf{n}}\dd S.
\label{ec:olla6_1}
\end{equation}
where the surface element on the boundary $\partial \Omega \equiv \mathcal{S}$ is written by means of the unit normal at each point of the surface, $\dd \mathbf{S} \equiv \hat{\mathbf{n}} \dd S$. We have also used the divergence theorem and the identity $\mathbf{v} \cross \curl \mathbf{A}  \vdot \hat{\mathbf{n}}= \mathbf{v} \vdot \curl \mathbf{A}  \cross \hat{\mathbf{n}}$.

There exist, indeed, dual spaces made of functions $\vb{v}$ such that $\vb{v} \vdot\curl \mathbf{A} \cross \hat{\mathbf{n}} = 0$ in the surface boundary $\partial \Omega$. Based on these function families, and using the previous notation for the functionals, we have
\begin{equation}
\langle \curl \mathbf{A}, \curl \mathbf{v} \rangle + \langle k \mathbf{A}, \mathbf{v} \rangle = \langle \mathbf{f}, \mathbf{v} \rangle
\label{ec:olla11_2}
\end{equation}
and the system of equations that comes out of Eq.~\ref{ec:olla11_2} in finite-dimensional function spaces is completed by imposing the specific boundary conditions for the unknown field $\vb{A}$.  The formulation Eq.~\ref{ec:olla11_2}, called ``weak formulation'' of the mathematical problem Eq.~\ref{ec:olla2_6}, is enough in our case because there are no currents \emph{through} the domain $\Omega$ (all currents stay inside it). 
More technical discourses about this variational form and boundary conditions for more general cases to solve Eq.~\ref{ec:olla2_6} can be found e.g.~in \cite{Hiptmair2008A-Robust-Maxwel,Olm2019On-a-general-im}.

\subsubsection{\label{subsec:2D version}2D calculations}

In this work we solve also the equivalent 2D problem. The reason is that 2D calculations can be compared directly with those from Ref.~\cite{lopezbruna20212Delectromagnet}, from which we have borrowed the plasma parameters and conductivity models. Therefore, the strategy is to check such models and profiles in 2D FEM calculations and then compare the latter with the corresponding 3D calculations based on exactly the same physical models and input parameters.

If, in Eq.~\ref{ec:olla2_0}, we make use of the electric field $\vb{E}$ instead of the vector potential, we have $\partial_t \curl \mathbf{B} = - \curl \curl \mathbf{E} =  \laplacian \mathbf{E}- \grad (\div \mathbf{E})$. As before, we neglect charge accumulation so that $\div \vb{E} \approx 0$. Recalling all possible current densities, which now include the coil currents, $\vb{J}_\mathrm{b}$, we write
$\laplacian \mathbf{E} = \mu_0 \partial_t (\vb{J}+\vb{J}_\mathrm{b}) + \mu_0 \partial_{tt} (\epsilon \mathbf{E})$. 
Using the Laplacian of a vector in cylindrical coordinates and considering, as it is pertinent for the 2D case, that the only component of the field is the azimuthal one, $E_\theta$, we are left with just one differential equation in the radial, $r$, and axial, $z$, coordinates,
\begin{equation}
\frac{1}{r}\partial_r (r \partial_r E_\theta ) + \partial_{zz}E_\theta - \frac{E_\theta}{r^2}  = \mu_0 \partial_t (J_\theta+J_{\mathrm{b}\theta}) + \mu_0 \partial_{tt}(\epsilon E_\theta),
%\label{ec:lido17_6}
\end{equation}
which is further simplified under our assumptions of purely harmonic response and the fact that, in the whole calculation domain, we can safely neglect displacement currents, 
\begin{equation}
\frac{1}{r}\partial_r (r \partial_r E_\theta ) + \partial_{zz}E_\theta - \left (\frac{1}{r^2} + \imath \omega \sigma \mu_0 \right ) E_\theta = \imath \omega \mu_0 J_{\mathrm{b}\theta}.
\label{ec:lido17_6}
\end{equation}
This can be re-written as a scalar equation for $E_\theta$ in the cartesian plane $(z,r)$, where the Laplacian is $\laplacian \equiv \partial_{zz}+\partial_{rr}$. Therefore we have
\begin{equation}
\laplacian E_\theta +\frac{1}{r} \partial_r E_\theta - \left ( \frac{1}{r^2} + \imath \omega \sigma \mu_0 \right ) E_\theta = \imath \omega \mu_0 J_{\mathrm{b}\theta};
\label{ec:lido90_2}
\end{equation}
or, defining $\mathcal{D}_\sigma E_\theta  \equiv  [ r^{-1}\partial_r  -(r^{-2}+ \imath \omega \sigma \mu_0 )] E_\theta$, as the differential equation $\laplacian E_\theta + \mathcal{D}_\sigma E_\theta  = \imath \omega \mu_0 J_{\mathrm{b}\theta}$. We do this simply to make the problem formally similar to the Poisson problem, for which the variational formulation is  well known. Let $v(x)$ be a space function in the domain of interest,  $x \in \Omega$. Since  $\nabla \vdot (v \nabla E_\theta) = \nabla v \vdot \nabla E_\theta + v \nabla^2 E_\theta$, we have
\begin{equation}
\int_\Omega \nabla^2 E_\theta(x) v(x) \dd x = - \int_\Omega \nabla v (x) \vdot \nabla E_\theta(x) \dd x - \int_{\partial \Omega} v(x) \nabla E_\theta(x) \vdot \dd \mathbf{S}
\label{ec:lido92_2}
\end{equation}
where the last integral concerns the flux directed outwardly from $\Omega$ through the boundary $\partial \Omega$. In the 2D case this is the integral along the 1-D contour of the domain. Choosing a family of functions that are null in this boundary, the variational problem 
becomes
\begin{equation}
-\langle \nabla E_\theta , \nabla v \rangle + \langle \mathcal{D}_\sigma E_\theta, v \rangle =  \imath \omega \mu_0  \langle J_{\mathrm{b}\theta}, v \rangle.
\label{ec:lido92_3}
\end{equation}
This is the variational problem to be solved in 2D via FEM. %, where nodal elements can be used for the scalar field $E_\theta$ in the plasma single domain. 

\subsection{\label{subsec:condiciones de contorno}Boundary conditions}

Ideally, a large calculation domain should be chosen where null fields can be safely imposed, to the desired approximation, as boundary values for the field. This, however, makes the problem considerably more difficult (surrounding materials and appropriate current densities in the RF coils must be included) and, especially, very intensive from the numerical viewpoint. At the present stage of development, it is simpler and practical  establishing a reduced calculation volume that excludes the RF coil and imposing, based on 2D calculations, reliable boundary conditions on its surface.

Boundary conditions based on 2D calculations can be easily justified in our case. From the experiments we know that the inductance of the void driver (without plasma) changes little, $\sim 1$\%, in presence of the plasma. As shown in \cite{lopezbruna20212Delectromagnet}, the calculations can reproduce this behaviour if the plasma conductivity decreases strongly with the RF itself. 
This means that the induced fields near the cylindrical boundary of the plasma remain basically unchanged during the discharges. With better reason, the induced fields in the close vicinity of the RF coil will be, to a good approximation, the same with or without plasma. At the same time, the elements that are relevant to calculate the vacuum field near the coils are axisymmetric; i.e., they are amenable to a 2D calculation. Therefore, we can extend the calculations to a larger domain where zero-field boundary conditions can be imposed due to the presence of metallic parts (except for the Faraday shield) and where the RF coil is included.

\begin{figure}[htbp] %  figure placement: here, top, bottom, or page
   \centering
   (a)\includegraphics[width=0.4\columnwidth]{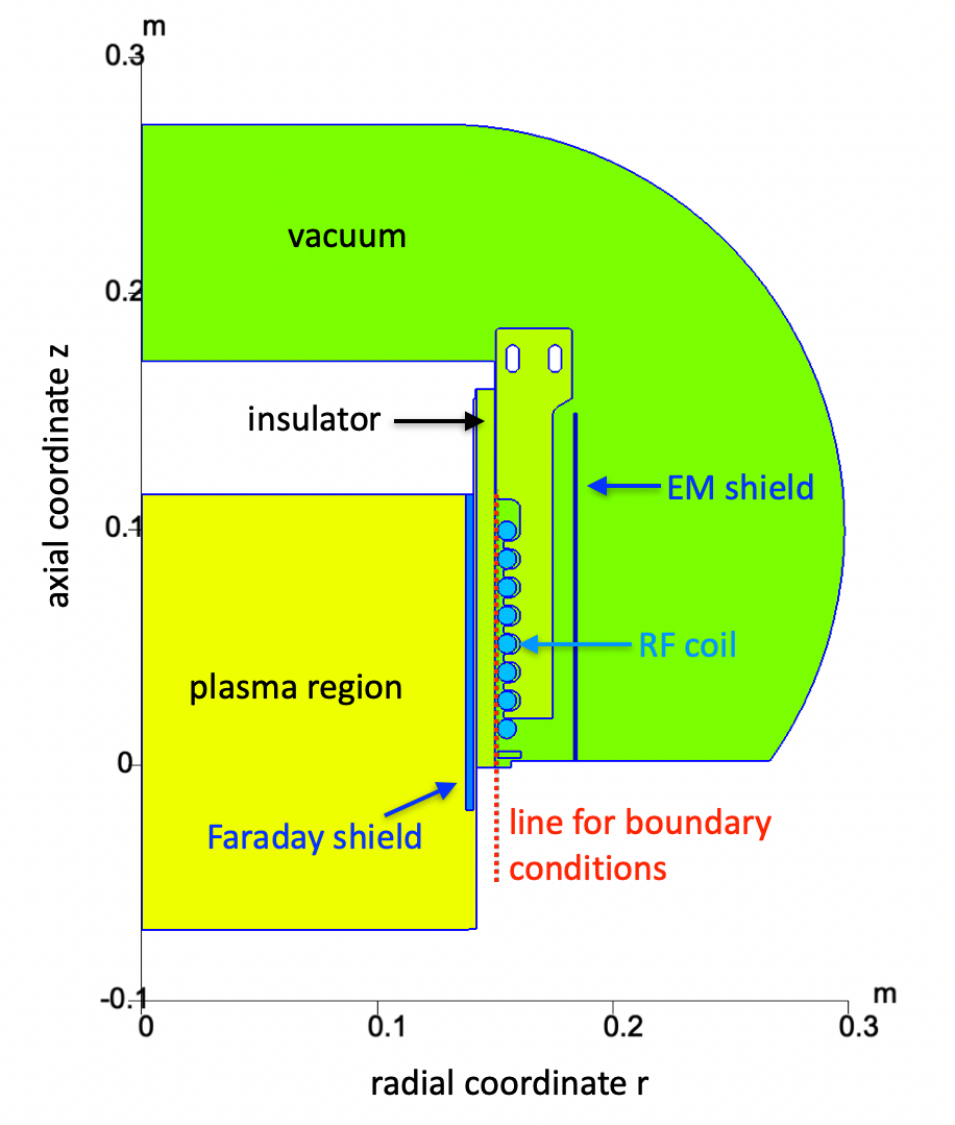}
   (b)\includegraphics[width=0.4\columnwidth]{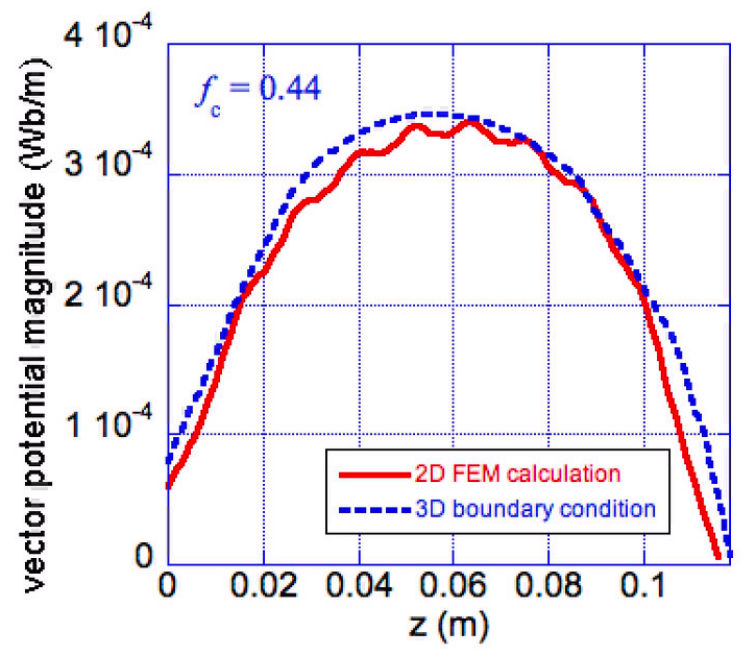} 
   \caption{(a) 2D domain from which the boundary field at the dotted red line are calculated, and (b) its comparison with the chosen boundary field in 3D calculations, shown in the $z$ coordinate of the 3D model.}
   \label{fig:contorno}
\end{figure}

Figure \ref{fig:contorno}a shows the 2D domain in which we have performed 2D calculations of the vacuum field. The labels indicate the different parts considered in the problem: plasma region (normally void), RF coil and electromagnetic shield. We also label the Faraday shield cylindrical wall, despite the fact that it is taken as ``transparent'' to the induced fields for these calculations. The evaluation of the field by the RF coil will be slightly overestimated. In order to quantify this omision we will make a comparison with experimental measurements of the driver inductance in vacuum.  We set null fields in the entire outer contour, which extends into the plasma expansion region. The calculations consist in setting a fixed plasma current $I_\mathrm{RF}$ in the RF coil and then obtaining the vacuum field along the red dashed line in figure \ref{fig:contorno}a, which indicates the boundary for the 3D problem in radial coordinate, about $1.4$ cm beyond the external lateral side of the Faraday shield.
 The azimuthal vector potential obtained from a 2D calculation along this line is shown in figure \ref{fig:contorno}b, where the effect of the nearby coils can be appreciated in small bumps.
 
  The magnetic flux through the RF-coil is easily obtained as the circulation of the vector potential. Considering that the RF coil is a set of eight single perfectly circular coils, we evaluate a vacuum driver inductance  $L_\mathrm{d}^\mathrm{vac} =  9.55$ $\mu$H. This value is in agreement with other calculations and, particularly,  measurements: $L_\mathrm{d}^\mathrm{vac} =  9.63$ $\mu$H \cite{jain2022investigation-o}. In figure \ref{fig:contorno}b we also plot the function adopted for 3D calculations. It is a modification of the theoretical vacuum solution that corresponds to the set of eight circular coils occupying the positions shown in figure \ref{fig:contorno}a. This theoretical function is proportional to $I_\mathrm{RF}$ and extends much longer axially. For this reason, we use smooth step functions at both extremes along $z$ to make it shape according to the numerical 2D solution \cite{lopezbruna20212Delectromagnet}. The maximum value is set via a reduction coefficient, $f_\mathrm{c} = 0.44$. As shown in figure \ref{fig:contorno}b, the comparison between calculated and ``manufactured'' functions is acceptable for our present purposes.

\subsection{\label{subsec:integracion} Numerical integration}

Building a numerical model from scratch to solve the present 3D problem is not practical unless one is interested in the development of numerical algorithms. On the other hand, there are many software suites with built-in solutions for the core calculations, like gridding and providing the FEM environment (finite element families, definition of domains and associated functions, matrix-inversion libraries etc.) We have chosen the numerical tools offered by the FEniCS project \cite{fenicsproject,Logg2012Automated-Solut,Langtangen2017Solving-PDEs-in} for the FEM calculations. FEniCS provides an appropriate language to manage integral forms \cite{Alnaes2012UFL} and a large set of tools to code FEM problems \cite{Logg2010DOLFIN,Logg2012DOLFIN}. The 3D mesh has been built with a separate software \cite{Geuzaine2009Gmsh:-A-3-D-fin} and then read from Python codes based on the FEniCS environment.

It is common knowledge that FEM problems dealing with the double-curl operator counsel using edge elements in order to preserve the tangential components of the unknown vector field. In solving Maxwell's equations, N\'ed\'elec elements are a typical choice. However, we have encountered that iterative solvers give unphysical solutions for $\vb{A}$ in many cases.  The large kernel of the double-curl operator lets gradient functions seep into the numerical solution, which is verified by noting that the magnetic field $\curl \vb{A}$ comes out well despite the unacceptable vector potential. In our case, however, ohmic dissipation is a very important numerical diagnostic. Since it involves directly $\vb{E} \sim \vb{A}$, we must obtain a correct vector potential.

  For one-domain (only plasma) calculations we can use nodal elements and iterative solvers by imposing the Coulomb gauged potential, $\div \vb{A} = 0$, from which we solve an equation involving the divergence operator; i.e., we write the variational formulation for equation $- \laplacian \vb{A} + k\vb{A} = 0$ instead of equation \ref{ec:olla2_6}. The Coulomb condition is justified in our case: the low frequency $\omega = 2\pi \cross 10^6$ rad/s gives a negligible contribution to the Lorenz condition, $\div \vb{A} = -\mu_0\epsilon_0 \partial_t V$, where $V$ is the electric potential. In our case $|\div \vb{A} | \sim 10^{-10}V$, a negligible term, especially when we are not considering capacitive processes.

We are generally interested in a problem with several calculation domains, e.g.~plasma and surrounding materials like the Faraday shield or the insulator. One difficulty associated to 3D calculations with domains of very different electrical properties is that edge elements are mandatory, unlike the one-domain cases referred to above, due to the boundaries between domains with largely different material properties where the boundary conditions of the unknown vector field must be preserved. In this case iterative solvers no longer converge towards acceptable physical solutions for the reasons indicated above.  The problem associated with the kernel of the double-curl operator might be alleviated or corrected applying specific techniques to solve the numerical system (e.g. \cite{Reynolds-Barredo2020A-novel-efficie}). This would imply intervening deeply in the numerical system, a task that is out of our possibilities. For this reason, the present work is based on calculations limited by the amount of memory available to manage the grid and the consequent matrices to be inverted via direct solvers. Iterative solvers, as mentioned, can be used in one-domain calculations using nodal elements; or, else, using edge elements with non-negligible conductivities.

% ======== MODELO GEOMƒTRICO ============

\section{\label{sec:cazuela}Driver geometrical model}

\begin{figure}[htbp] %  figure placement: here, top, bottom, or page
   \centering
   \includegraphics[width=0.45\columnwidth]{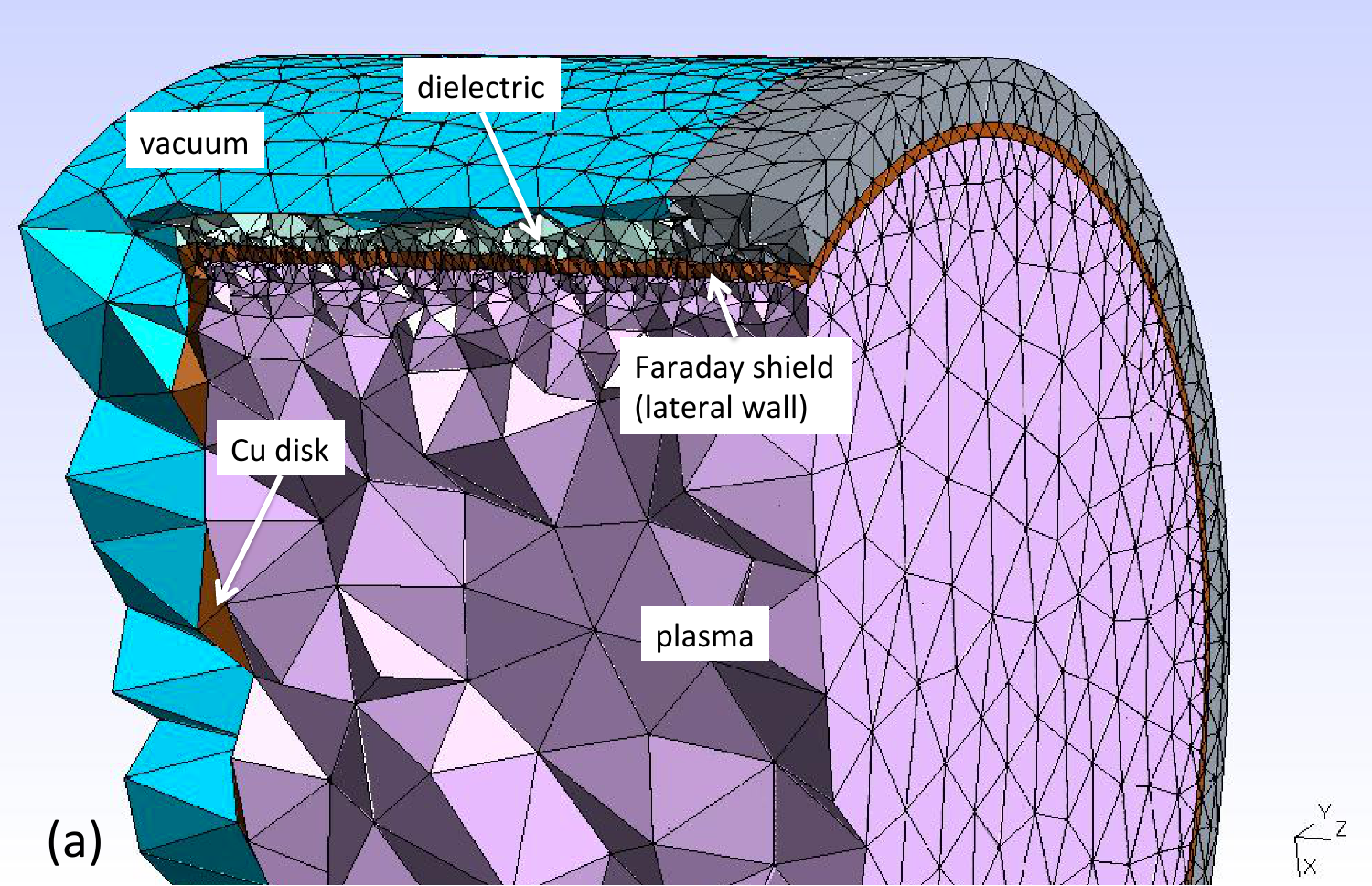} 
   \includegraphics[width=0.45\columnwidth]{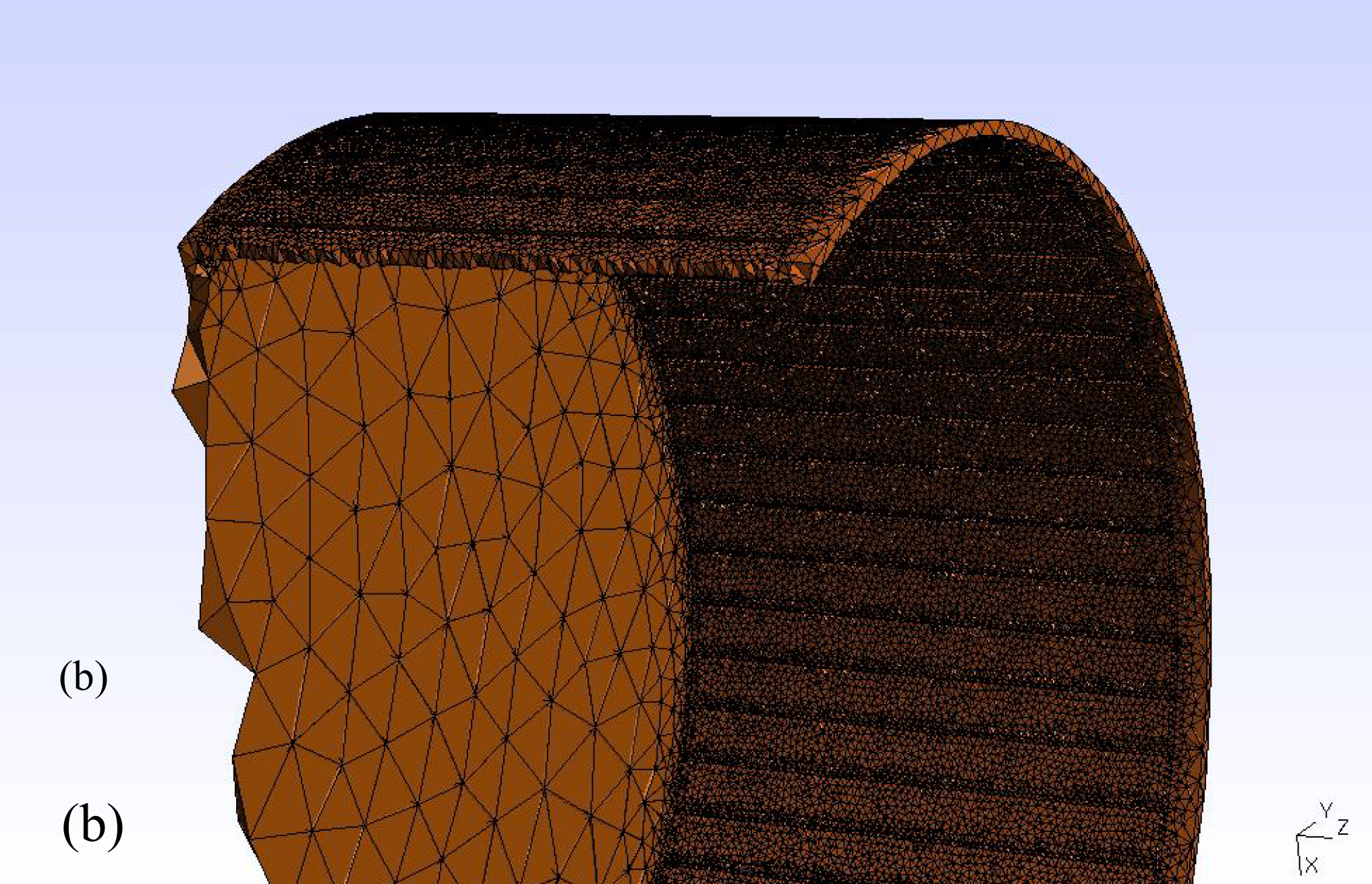} 
   \caption{(a) Cut of the driver model mesh showing the different sub-domains of the 3D calculations. (b) Isolated view of the Faraday shield in the model, which includes the copper back disk and cylindrical lateral wall. }
   \label{fig:cut_driver_model}
\end{figure}

The driver is described numerically using a simplified model of the different parts that compose any of the eight real drivers of SPIDER, see figure \ref{fig:real driver}. Since the calculations do not include the RF coil, we have modeled the driver considering the main metallic parts inside a cylinder whose radius almost reaches the coil. The geometrical model of the driver consists of subdomains where different material properties and mesh sizes can be assigned, as illustrated in figure \ref{fig:cut_driver_model}a. The core part is the plasma region, which is open to the plasma expansion region (towards the right in the figure) but limited by a copper disk at the opposite side and by the cylindrical lateral wall of the Faraday shield. The latter is surrounded by an insulator, which finishes with a steel ring on the plasma expansion region side. The ring shape is due to the limited calculation domain, as this stainless steel region is simply a part of the planar support structure where all the drivers are inserted to face the expansion region: it covers the bottom part of the Faraday shield cylindrical wall in figure \ref{fig:real driver} as a prolongation of the insulator. The different elements in figure \ref{fig:cut_driver_model}a can be understood also in view of the 2D cut shown in figure \ref{fig:contorno}a. 
%Except for the steel ring, the ensemble of parts is surrounded by vacuum.
 Table \ref{T:calefactor} shows the ranges in radial and axial coordinates occupied by the different regions of figure \ref{fig:cut_driver_model}a.

\begin{table}
\centering
\caption{Range of coordinate values $r = \sqrt{x^2+y^2}$ and $z$ for the different parts of the model driver shown in figure \ref{fig:cut_driver_model}a.}
\begin{tabular}{lccl}
Part   & $r$ (cm) & $z$ (cm) & NOTE \\
\hline\hline
plasma & $0 < r < 13.725$ & $0 < z < 15.66$ & Except shield gaps  \\
\hline
\multirow{2}{*}{shield}  & $13.725 < r < 14.055$ & $0 < z < 15.66$ & Lateral wall  \\
   & $0 < r < 14.055$ & $-0.6 < z < 0$ & Back Cu disk \\
\hline
dielectric  & $14.055 < r < 14.955$ & $0 < z < 14.86$ & Alumina cylinder \\
\hline
Steel & $14.055 < r < 15.455$ & $14.86 < z < 15.66$  & \\
\hline
\multirow{2}{*}{vacuum}  &  $0 < r < 15.455$ & $-3.6 < z < -0.6$  & Vacuum disk\\
 & $14.955 < r < 15.455$ & $0 < z < 14.86$  & Vacuum cylinder
\end{tabular}
\label{T:calefactor}
\end{table}

\begin{figure}[htbp] %  figure placement: here, top, bottom, or page
   \centering
   \includegraphics[width=0.73\columnwidth]{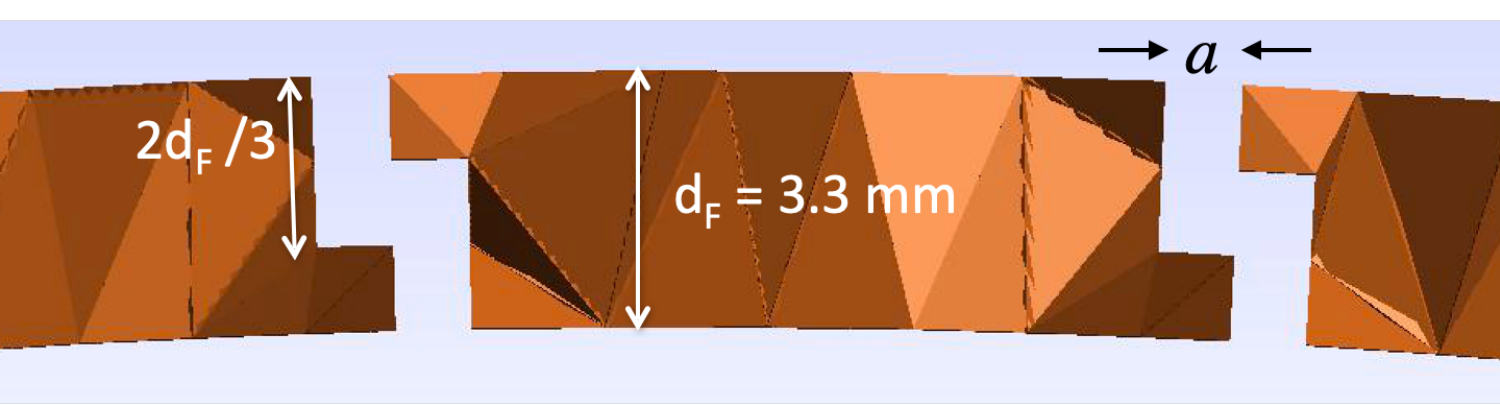}
   \caption{Detail of the section of the Faraday shield lateral wall on the $z$-plane, where $d_\mathrm{F}$ is its thickness in radius.}
   \label{fig:cut_slits}
\end{figure}

 The Faraday shield structure merits particular attention. Part of the need for 3D calculations is due to the breaking of axisymmetry in the Faraday shield lateral wall, which is intended to isolate the plasma and protect the dielectric material. To allow for the penetration of the RF field into the plasma region, the shield lateral wall has slits along the axial (to avoid strong azimuthal currents) direction. Figure \ref{fig:cut_driver_model}b  shows the model Faraday shield, where the slits between some of the 80 evenly-spaced ``roof-tiles'' can be seen. The mesh is particularly fine in this lateral wall, a necessary condition to resolve the fields inside the highly conducting copper, and becomes coarser in proportion to the inverse of the distance to it up to a prescribed size, as can be appreciated in the back disk of the shield. The mesh has been constructed with the Gmsh software \cite{Geuzaine2009Gmsh:-A-3-D-fin}, with which we have also produced figure \ref{fig:cut_driver_model}. %The minimum tetrahedron size is set so as to limit the number of elements located by the surfaces of the Faraday shield lateral wall. The maximum size can also be controlled in order to provide enough accuracy for the calculation in the plasma part of the domain.
Figure \ref{fig:cut_slits} presents part of a section of the Faraday shield with the $z$-plane, where the shape of the slits can be clearly seen. Taking the exterior circumference of the lateral wall section, the openings associated to the slits occupy approximately an $8.5$\% of the circumference. We shall refer to this percentage as clearance.

\begin{figure}[htp] %  figure placement: here, top, bottom, or page
   \centering
   \includegraphics[width=0.43\columnwidth]{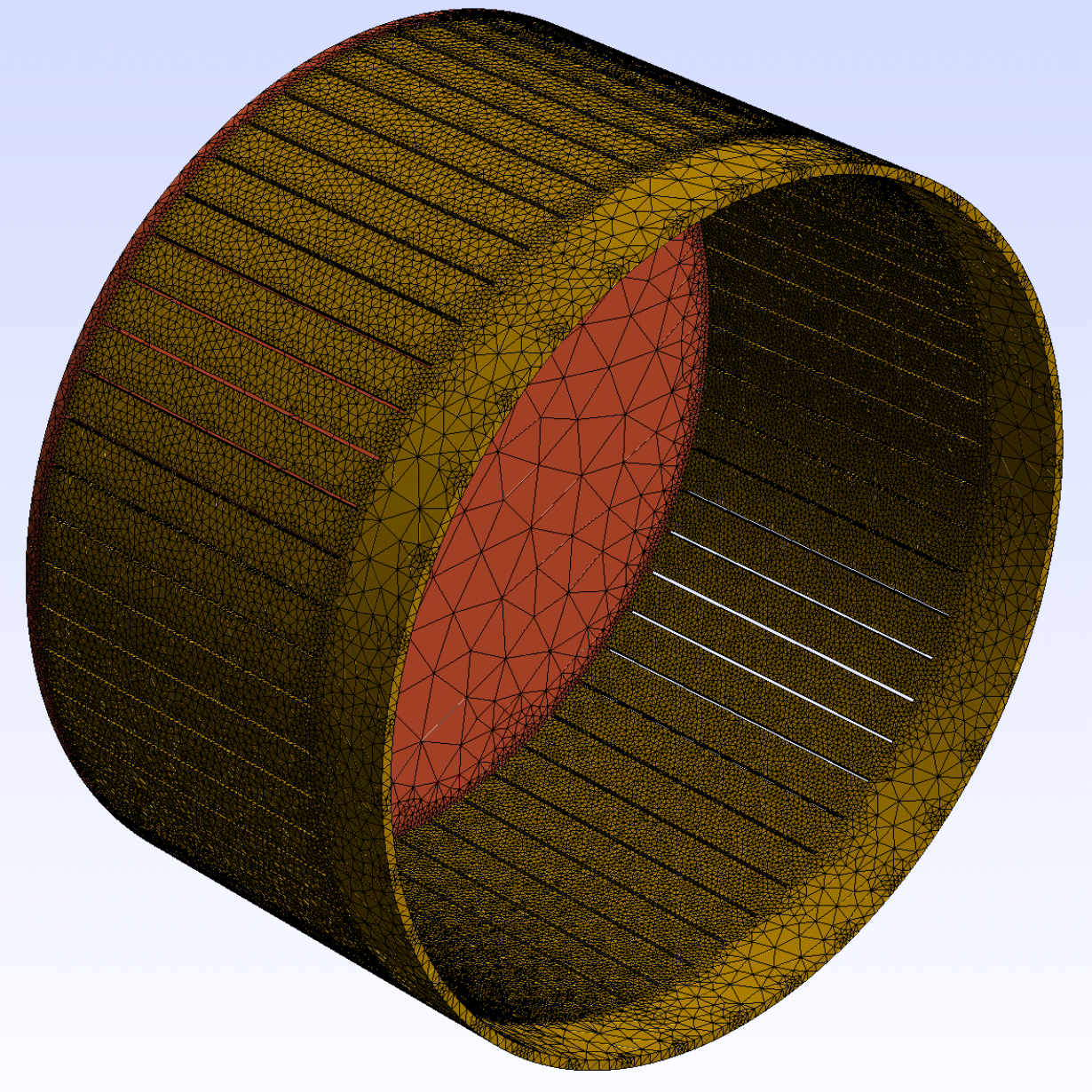}
   \caption{Simplified model of Faraday-shield with straight slits.}
   \label{fig:FaradayShieldStraight}
\end{figure}

For certain studies it is convenient using a simplified lateral wall, where the slits are straight in the radial direction. Such simplified lateral wall can be seen in Figure \ref{fig:FaradayShieldStraight}, %similar to figure \ref{fig:cut_driver_model},
 where the apertures, due to the slits being straight, can be appreciated. Here we have eliminated the steel disk, a domain that is now converted to vacuum, making instead the corresponding portion of the lateral wall continuous. In this simpler geometry we can easily define two control parameters related with the geometry of the lateral wall: the ``clearance ratio'', $\Delta_\mathrm{s}$, and the number of tiles, $N_\mathrm{T}$, coincident with the number of slits. The former is defined as follows. Let $2\pi r$ be the length of the outermost circumference of the lateral wall cross-section and let us consider the length $L$ that corresponds to the slits. If each slit has a width $a$, then $L=aN_\mathrm{T}$. Now we simply define the ratio $\Delta_\mathrm{s} = L/2\pi r$. Since the slits go along most of the entire length of the Faraday shield in the axial direction, this ratio gives a rough estimate of the percentage of the lateral wall area that is ``open to the plasma''. For instance, a value $\Delta_\mathrm{s} = 0.1$ roughly indicates a 10\% clearance.

Note that the same definitions can be applied to the more realistic lateral wall in figure \ref{fig:cut_slits}: obviously the number of tiles, but also $\Delta_\mathrm{s}$. The difference is that characterising the cross-section of this kind of wall would require more parameters. We have preferred to set the simpler Faraday shield with straight slits for scanning the effects of the clearance ratio, $\Delta_\mathrm{s}$, and the number of tiles, $N_\mathrm{T}$. The results are presented in next section.

\section{\label{sec:resultados}Calculation results}

\begin{figure}[ht] %  figure placement: here, top, bottom, or page
   \centering
   \includegraphics[width=0.73\columnwidth]{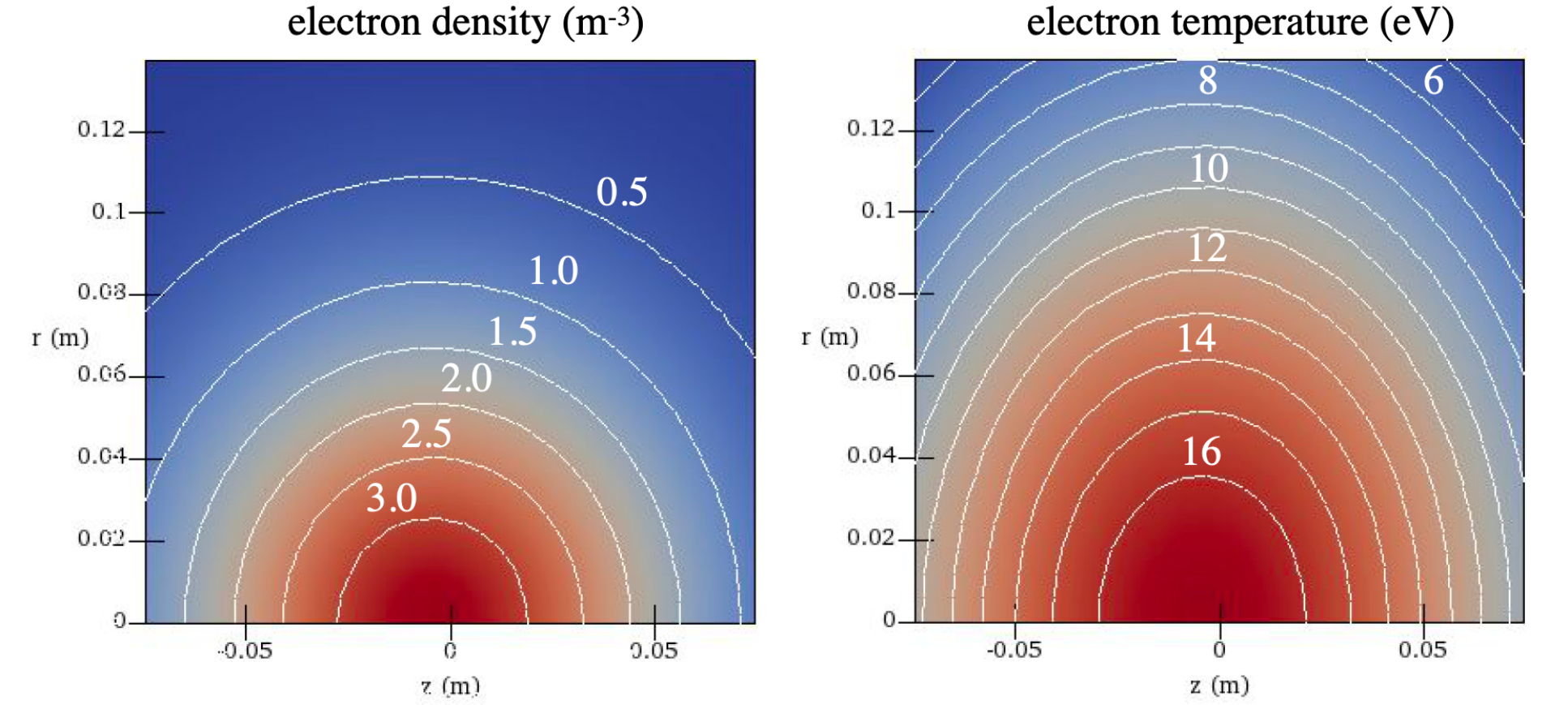}
   \caption{Model distributions of electron density and temperature in the axial--radial plane of a SPIDER plasma. Net power: $50$ kW; Hydrogen gas pressure: $0.34$ Pa; filter magnetic field from plasma-grid current $I_\mathrm{PG}=2.6$ kA. The axial coordinate, $z$, is centered in the middle of the axial direction of the driver.}
   \label{fig:densi_tempe}
\end{figure}

\subsection{\label{subsec:compara 2D}Comparison with 2D calculations}

The calculations that follow use input data from characteristic plasmas of the S16 campaign of SPIDER (y.~2020), as described in Ref.~\cite{lopezbruna20212Delectromagnet} where 2D electromagnetic calculations were done with a finite element code. 
In order to confront these results with 3D calculations we must use the same two kinds of plasma: without and with static filter magnetic field, both under $50$ kW nominal power per driver. Figure \ref{fig:densi_tempe} presents contour maps of the smooth 2D functions chosen to represent the experimental data of the plasma with filter field, for which the peak electron density and temperature values inside the driver are, respectively, $3.4 \cross 10^{18}$ m$^{-3}$ and $17$ eV. The plasma without filter magnetic field has smaller peak electron density, $1.2 \cross 10^{18}$ m$^{-3}$, and almost flat electron temperature, $11$ eV.

We start by comparing the results obtained on a same grid of $N_z \cross N_r$ rectangular elements using either the ``finite differences'' (FD) code of \cite{lopezbruna20212Delectromagnet} or the ``finite elements'' (FE) version in 2D. The visual comparison of the results is very satisfactory, and integral values like the absorbed ohmic power or net plasma current are the same at all practical respects. In order to better quantify the differences, we use the Euclidean norm of the solution functions $E_\theta (z,r)$ taken as a vector of dimension $N_z \cross N_r$; i.e., if we have the discrete numerical solution $E_j =E_\theta (z_j,r_j)$ in each of the $j$-th calculation nodes, we define
$$
|E| = \sqrt{\sum_{j=1}^{N_z \cross N_r} E_j^2}.
$$
In particular, we calculate this norm for the difference between the solution obtained by each method, $\Delta E \equiv | E^\mathrm{FE}-E^\mathrm{DE} |$. 
 Rounding the results of the induced electric field to $1$ V/m, we obtain $\Delta E = 11$ V/m. We get an idea of its meaning by calculating the separate norms  $|E^\mathrm{FD}| = 25728 \approx |E^\mathrm{FE}| = 25725$ V/m. Note that the numerical approach is completely different in both cases, and also that the calculations involve iterative procedures to find the plasma conductivity and the RF magnetic field \cite{Jain2018Improved-Method,lopezbruna20212Delectromagnet}. The two methods are equivalent as calculation tools.

\begin{figure}[htbp] %  figure placement: here, top, bottom, or page
   \centering
   \includegraphics[width=0.33\columnwidth]{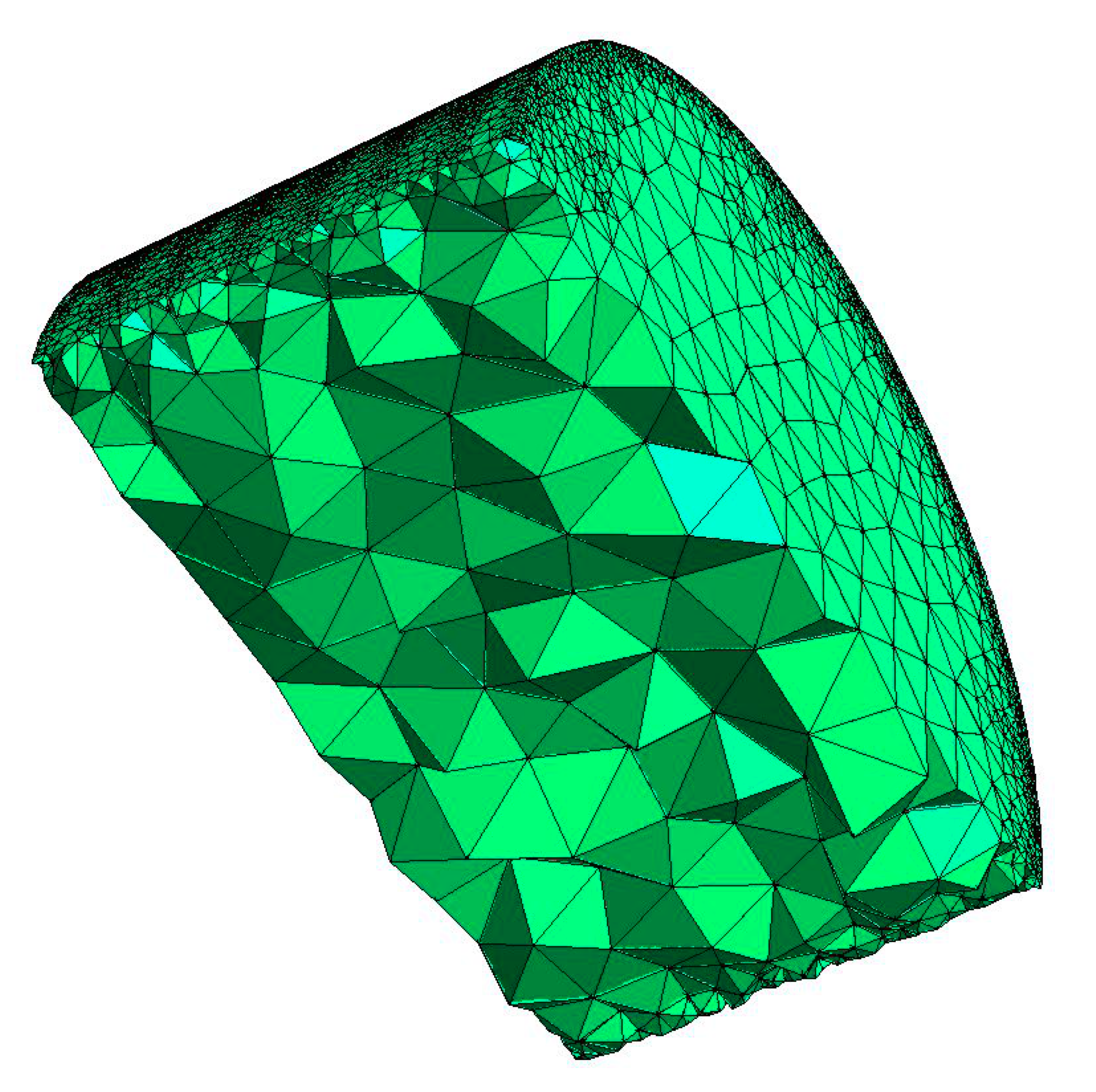}
   \includegraphics[width=0.4\columnwidth]{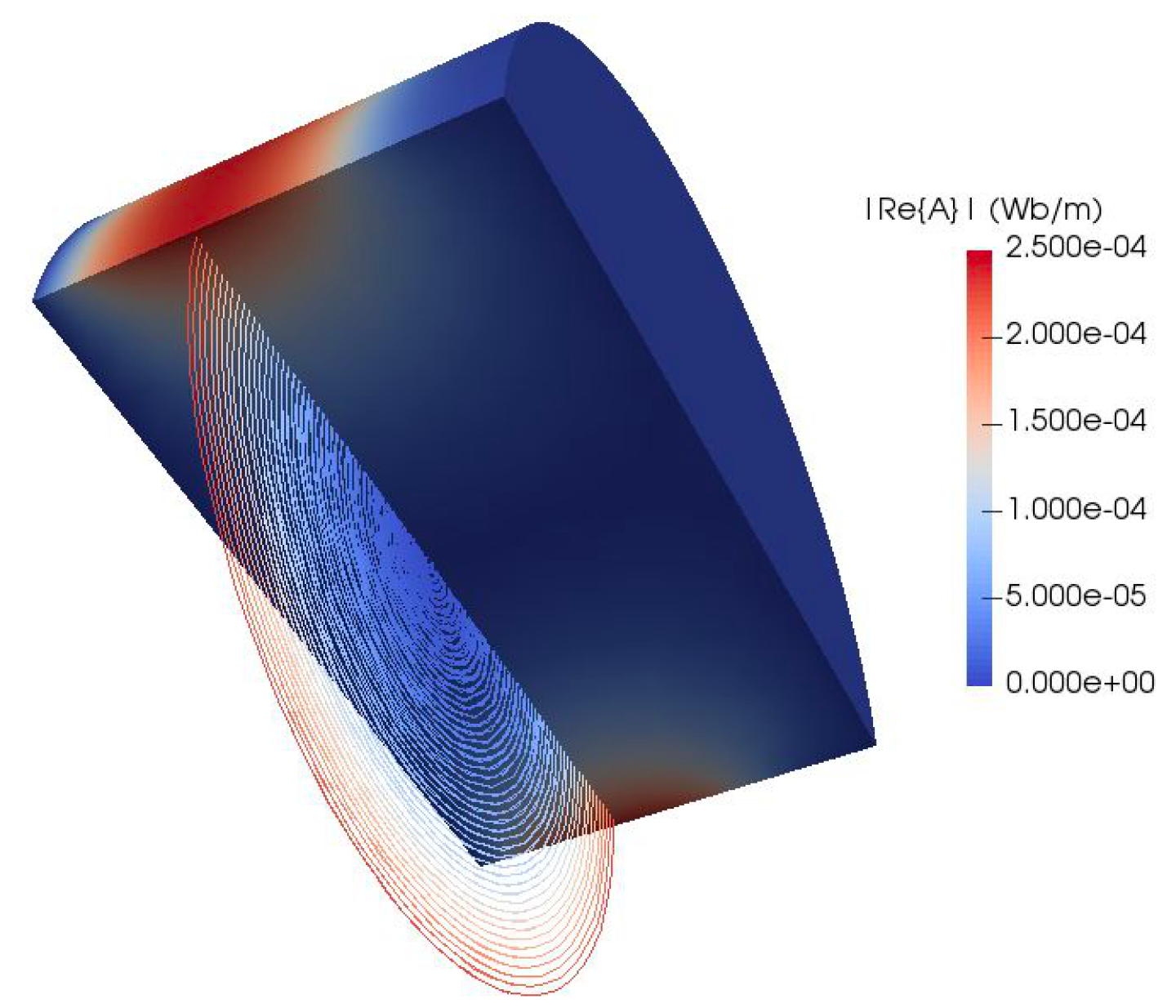} 
   \caption{Left: Mesh for 3D calculations with only the plasma domain. Right: Magnitude of the vector potential (real part) and field lines integrated from a line passing through a vertical axis near the maximum.}
   \label{fig:olla_2}
\end{figure}

 Next we check that the 2D solution remains in 3D calculations based on the variational formulation for the magnetic vector potential, equation \ref{ec:olla11_2}. %The nature of the problem counsels using so called ``edge finite elements''.  When only the plasma domain is considered, as is the case now, iterative solvers allow for efficient 3D parallel computations.
  The calculations are done on the same plasmas as above and using all the elements to make the cases equivalent, like boundary conditions, electrical parameters etc.  The results, obviously, depend on the chosen meshes if one wants to find a limit for accuracy. This is not critical to establish the similarity of results among 2D and 3D calculations.

  A good way to check for the axi-symmetry of the solution is to compare the axial component of $\vb{A}$ with any of the perpendicular components. We have done this exercise using the mesh shown in figure \ref{fig:olla_2} (left), where we can appreciate the refinement near the cylindrical boundary, reaching tetrahedra with edge sizes $1$--$3$ mm (compare with the cylinder radius, $149$ mm). We have set a space function of degree-3 polynomials in each cell of this mesh. The  integration of several field lines with the software ``ParaView'' \cite{ahrens2005paraview} illustrates the axi-symmetry of the system, see Figure \ref{fig:olla_2} (right). Looking at the numbers we find that, in most of the domain, the axial components are smaller than $1/1000$ the perpendicular ones in magnitude, with oscillating sign. Only in a few spots near the junction between the cylindrical and flat boundary surfaces this factor approaches $1/100$. We can attribute these local non-null axial components of $\vb{A}$ to the conversion of order 2 functions in the FEM calculations to linear functions in the visualization software. The obtained 3D solution field gives, in practice, the same integrated power and the same field magnitude than the 2D version. We illustrate this in Table \ref{T:olla_caz053_cr_olla028}, where two calculations with coarser grids ($24 \cross 26$ nodes in 2D; and $760$ nodes in 3D) are added to show that these simple calculations do not require fine grids in practice. 

\begin{table}[htbp]
\centering
\caption{Comparison of 2D and 3D FEM calculations on a SPIDER plasma using equivalent inputs (conductivity model, plasma profiles and electrical parameters). The outputs shown are the plasma absorbed power and the maximum values of the induced electric field.}
\label{T:olla_caz053_cr_olla028}
\begin{tabular}{lcccc}
geometry &   $P_\Omega$  & $\max \Im\{E_\theta\}$ & $\max \Re\{E_\theta\}$  \\
 & (kW) & (V/m) & (V/m) \\
\hline
\hline
\texttt{2D} ($24 \cross 26=624$ nodes)   & $18.97$ & $1503.8$ & $138.3$ \\
\texttt{2D} ($65 \cross 60=3900$ nodes)   & $18.97$ & $1503.8$ & $138.4$ \\
\texttt{3D} (17910 nodes)  & $18.97$ & $1504.2$ & $138.5$  \\
\texttt{3D} (760 nodes)  & $18.92$ & $1503.6$ & $137.6$  \\
\end{tabular}
\end{table}

With these comparisons we conclude that the 2D electromagnetic calculations performed so far for the drivers of SPIDER \cite{lopezbruna20212Delectromagnet} have been transported to 3D. This is the starting point to add intrinsically 3D features.

\subsection{\label{subsec:rendijas}Dissipation according to the geometry of the Faraday shield}

Starting from the one-domain (plasma) 3D calculations presented above, the different parts of the driver shown in figure \ref{fig:cut_driver_model} have been added step by step. For fixed boundary conditions at the surface of the model driver, the inclusion of metallic parts adds the corresponding dissipation. Thus, the net absorbed power (plasma and metallic parts) approaches from below the nominal power of the corresponding discharges as the metallic parts are added. As will be shown later, the dissipation is dominated by the Faraday-shield lateral wall.

An important element in the construction of an RF plasma source for neutral-beam heating is the Faraday shield. Its lateral wall must have slits to allow for the penetration of the magnetic field induction, but the absorbed power is sensitive to the geometric details. In order to simplify the study, we present here several calculations where the model lateral wall has been generated with straight slits (figure \ref{fig:FaradayShieldStraight}).

\begin{figure}[htbp] %  figure placement: here, top, bottom, or page
   \centering
   \includegraphics[width=0.45\columnwidth]{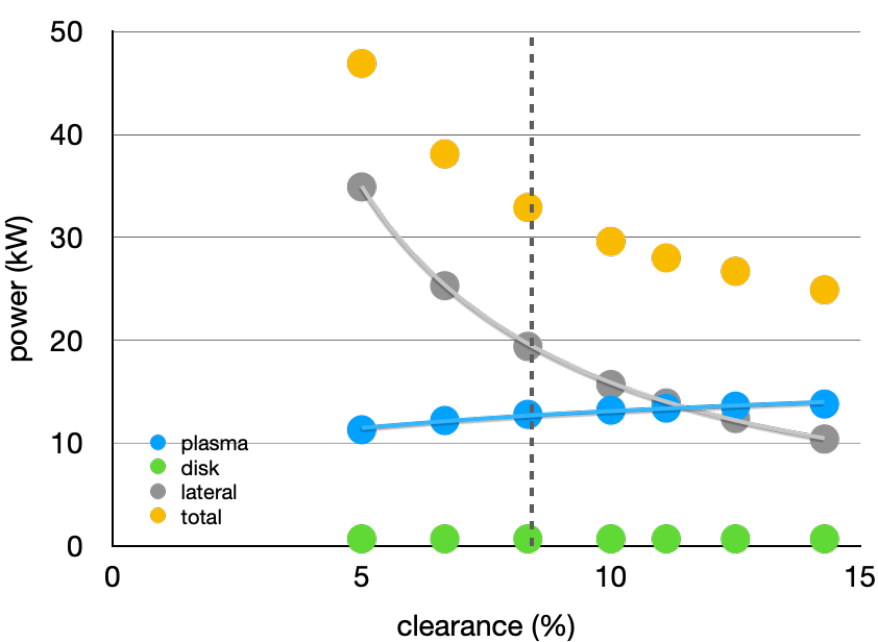} 
   \includegraphics[width=0.45\columnwidth]{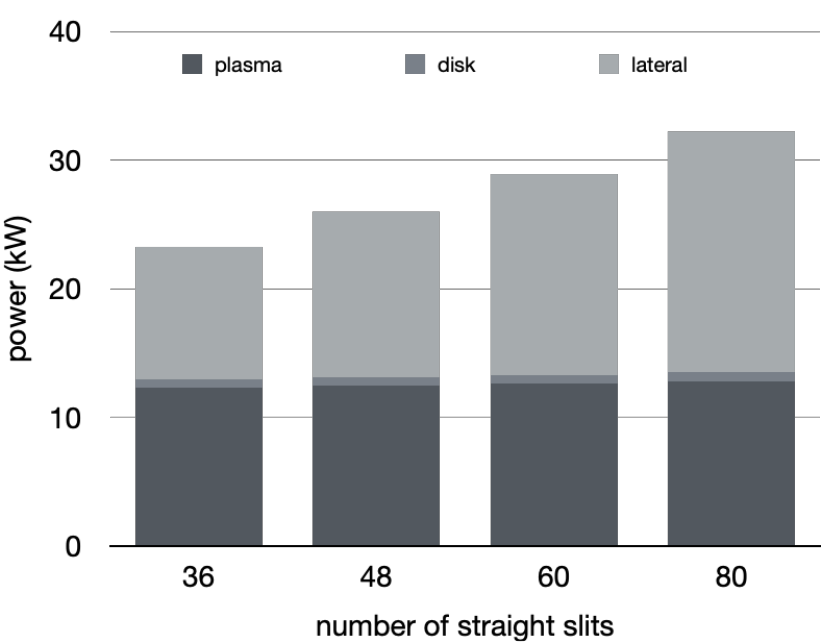}
\caption{Power share scans using a model Faraday shield with straight slits and no other metallic parts. Left: scan of the clearance using a shield with $N_\mathrm{T}=80$ straight slits. Right: scan on the number of slits for a fixed clearance ratio $\Delta_\mathrm{s} = 0.085$ (vertical line in left figure).}
   \label{fig:occupation_degree}
\end{figure}

In figure \ref{fig:occupation_degree} (left) we can see that a smaller clearance affects the absorbed power both in the shield (increases) and in the plasma (decreases).  This is expected because the field lines enter with more difficulty the plasma region when the clearance  diminishes; or, alternatively, the field sticks to the configuration that, compatible with the boundary conditions, permits a stronger opposition to the driving field of the RF coil. This is evident when one approaches the extreme cases: a clearance ratio that tends to zero would favour dissipating all the power in the Faraday shield, while a tendency to a 100\% clearance would give a dissipated power that tends to zero in the shield. In general, using a typically high number of slits, the role of the Faraday shield in the ohmic power losses is prominent in this type of plasma source, as also found elsewhere \cite{rauner2022impact-of-inter}.

The number of slits changes the power absorption for a fixed clearance ratio, as shown in figure \ref{fig:occupation_degree} (right). The power share is approximately the same in the different parts when the number of tiles is changed at a fixed $8.5$\% clearance; except for the Faraday shield lateral wall, where the dissipation increases with the number of tiles (or slits, equivalently). Eddy currents are generated around the internal faces of the slits. The larger its number, the larger the dissipation for a given clearance.

The real cylindrical wall of the Faraday shield has a different structure. There are 80 gaps with a different section, see figure \ref{fig:cut_slits}, as the roof-tiles are designed to protect the dielectric casing. For a fixed clearance, the dissipated power is slightly larger ($\sim 10$\%), both in plasma and lateral wall, than in the case of straight slits. Indeed, if the clearance and number of slits is fixed, the shape of the slits has little effect on the power share (also assuming a same plasma conductivity) \cite{wang2022design-optimiza}. 
  Therefore, the change of slit section has no importance for the trends found in the scans shown so far, but the comparison with experimental measurements must be done with the more realistic model.

% Drawings obtained with the ParaView software \cite{ahrens2005paraview}.

% As mentioned earlier, the mesh has been refined near the cylindrical metallic wall of the Faraday shield. This is obliged by the very large conductivity of the copper, $\sigma_\mathrm{Cu} \gtrsim 3\cross 10^{7}$ S/m at the temperatures of interest, and the corresponding tiny skin depth, $\sim 0.1$ mm at the relevant RF. This requires $\sim 10^6$ cells and order-two function families.

\subsection{\label{subsec:temperatura cobre}Power share with plasma}

3D calculations in \emph{one} domain have been used in Sec.~\ref{subsec:compara 2D} to assess that the equations and conductivity models previously used in 2D are available in 3D. However, the interest of 3D calculations relies on the possibility of studying the power share in different parts of the driver ---very importantly in the Faraday shield lateral wall--- and using more consistent conductivity models. This last aspect is left for a future work because it is an active field of research that requires dedicated studies (see, e.g. \cite{chen2006nonlinear-effec}). Here we keep on using the conductivity model used in 2D calculations \cite{Jain2018Improved-Method,Recchia2021studies-on-powe,zagorski20222-d-fluid-model,lopezbruna20212Delectromagnet,zagorski20232d-simulations} because they give acceptable values of deposited power in the plasma, i.e., whatever a correct model for the conductivity is, it should give values on the same order of those yielded by the models in use.

\begin{figure}[htbp] %  figure placement: here, top, bottom, or page
   \centering
   \includegraphics[width=0.4\columnwidth]{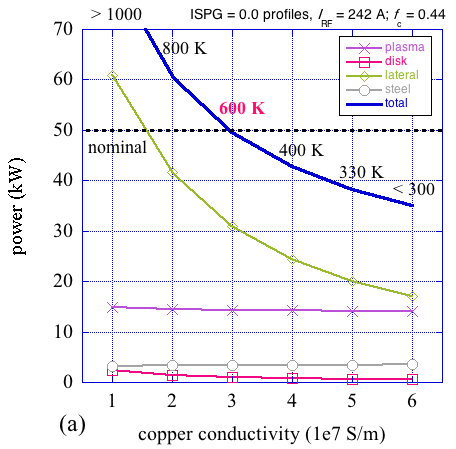} 
   \includegraphics[width=0.4\columnwidth]{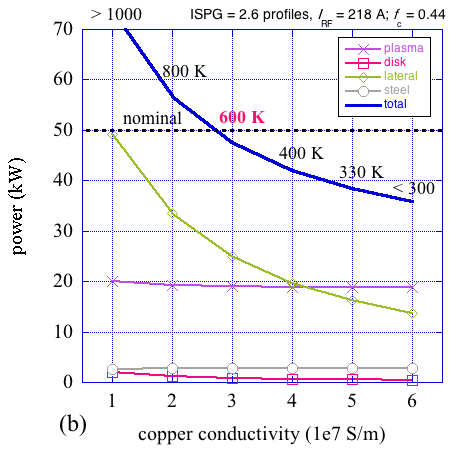}
   \caption{Power absorbed by the different parts considered in the calculations (see labels for each symbol) depending on the conductivity assigned to the copper parts. The temperatures associated to each conductivity value are indicated above the thick blue line, which represents the net absorbed power. The value $600$ K is highlighted (magenta) as a limit for electro-deposited copper. The calculations are done with experimental plasma parameters taken from discharges (a) without and (b) with filter static magnetic field. The nominal power is indicated with a dotted line to provide a reference. Other continuous lines are linear interpolations to guide the eye.}
   \label{fig:power_share_scan}
\end{figure}

Figure \ref{fig:power_share_scan} shows the power absorbed by the different parts accounted for in the calculation domain (see figure \ref{fig:cut_driver_model}) for two fixed distributions of plasma density and temperature, which correspond to discharges (a) without and (b) with filter magnetic field, depending on the conductivity of the copper parts. The power absorption in the Faraday shield (lateral wall and back disk) is obviously sensitive to the conductivity of copper, in turn dependent on its temperature. Since the shield is made of electro-deposited copper, the temperature should not surpass around 600 K (highlighted in the graph) in order to preserve its electro-mechanical properties. In any case, the power dissipated in the Faraday shield dominates the losses owing to the large power absorbed in its lateral wall. According to the figures, the cooler the copper parts, the more efficient the driver is. Note, however, that considering one only copper conductivity value for the estimates of figure \ref{fig:power_share_scan} is a simplified exercise in the sense that it does not consider the likely inhomogeneous RF power density distribution in the Faraday shield; and therefore, despite the water cooling protection and the good thermal conductivity of copper, there might be hot spots and cooler parts causing a feedback in dissipation. These considerations are left for ongoing studies.

The two panels in figure \ref{fig:power_share_scan} suggest $\sigma_\mathrm{Cu} \gtrsim 4 \cross 10^7$ S/m as acceptable values of the copper conductivities in these calculations: they correspond to admissible temperatures for the Faraday shield, while the resulting estimated net power, also considering other losses estimated on the order of several kW,  approaches the nominal power per driver, $\approx 50$ kW. Therefore, next calculations are based on $\sigma_\mathrm{Cu} = 4 \cross 10^7$ S/m for the copper parts. On the other hand, the absorbed powers depend on the amplitude of the RF-coil current, $I_\mathrm{RF}$. The values used to obtain the figures \ref{fig:power_share_scan} (a) and (b) are based on the respective values obtained with  previous studies for the SPIDER device \cite{jain2022investigation-o}, see indications on top of the figures. But the present calculations allow for an independent estimate of $I_\mathrm{RF}$ guided by the net absorbed power.

\begin{figure}[htbp] %  figure placement: here, top, bottom, or page
   \centering
   \includegraphics[width=0.85\columnwidth]{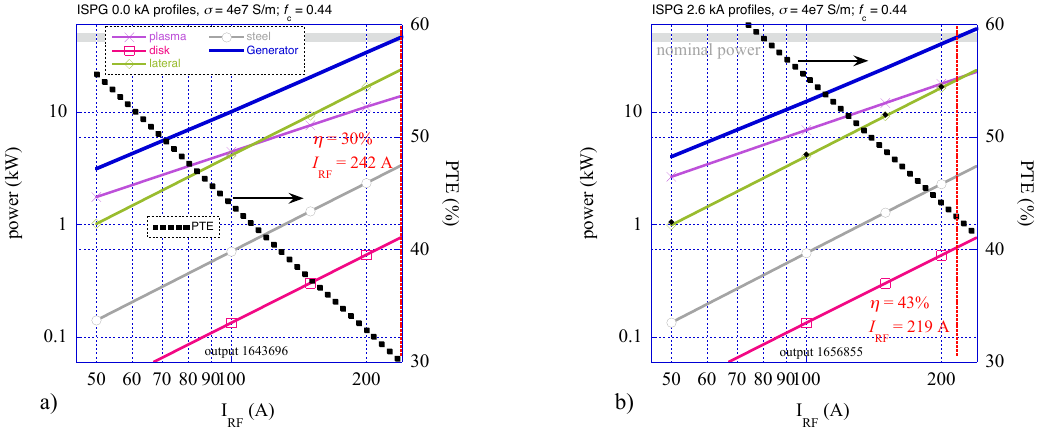}
   \caption{Scan on the RF-coil current, $I_\mathrm{RF}$, to obtain the power share (left ordinate) and consequent plasma transfer efficiency (right ordinate) in calculations with $\sigma_\mathrm{Cu} = 4 \cross 10^7$ S/m. The values correspond to plasma discharges without (a) or with (b) filter magnetic field. A horizontal grey band marks the nominal power of the discharges, $\approx 50$ kW. Absorbed powers obtained for the Faraday shield lateral wall in the absence of plasma are indicated with black diamonds in (b).}
   \label{fig:LIBERTY34_1}
\end{figure}

Figure \ref{fig:LIBERTY34_1} shows, in logarithmic scale, the result of scanning the power share as a function of $I_\mathrm{RF}$.  The scans are not realistic in physical terms because the plasma profiles are kept fixed while, in reality, they would be very different, or even non-existent, when the current is changed in such a wide range. On the other hand, this exercise is necessary in order to check the scalings of the power depending on the physical domain (plasma, shield, steel...) and, moreover, there must exist a range of $I_\mathrm{RF}$ for which the imposed (experimental) plasmas are realistic. This should correspond to a total dissipation near the generator nominal power per driver, $50$ kW in our case.

 The power transfer efficiency (PTE) is the ratio between plasma absorbed power and the net power delivered to the driver. We can define the PTE in terms of the calculated values as plasma absorbed power over some estimate of the net absorbed power, which can be obtained  complementing 3D calculations as follows. There are two main passive metallic parts, both beyond the 3D domain of our calculations, that can contribute appreciably to the dissipated power: the RF coil and the electromagnetic shield.  We use the effective resistance that corresponds to the RF coil according to the measurements, about $R_\mathrm{RF}=0.12$ $\Omega$. Therefore we add a contribution $R_\mathrm{RF}I_\mathrm{RF}^2$ to the losses. The effective resistance that corresponds to the electromagnetic shield can be estimated using 2D calculations in the geometry of figure \ref{fig:contorno}, where the dissipation in the electromagnetic shield is found to respond as $R_\mathrm{EM}I_\mathrm{RF}^2$ for a constant effective resistance $R_\mathrm{EM}$ associated to the electromagnetic shield. We find $R_\mathrm{EM} = 0.04$ $\Omega$. We add these two contributions to the 3D calculated dissipation in the plasma, Faraday shield and steel parts. Thick blue lines in figure \ref{fig:LIBERTY34_1} represent the resulting net dissipation from which we obtain the PTE represented with broken lines.
 
The RF-coil current amplitude and the efficiency that would justify the experimental nominal power is labelled and marked  with vertical red dotted lines in figure \ref{fig:LIBERTY34_1}. The values of coil current are nearly identical to those of previous studies for the corresponding plasmas (see indications on top of figures \ref{fig:power_share_scan}). Note, however, that this striking similarity of $I_\mathrm{RF}$ values is a coincidence in view of figure \ref{fig:power_share_scan} because we know that different values of $\sigma_\mathrm{Cu}$ will produce some variation of these estimates. Roughly speaking, we can conceive a variation of the PTE around a 10\% for similar variations of the conductivity in the copper parts. In any case, we can assert that the present 3D calculations provide  reasonable estimates of the driver resistance, with or without plasma.
  
Since power-law fits to the calculated dissipation in the metallic parts yield quite precisely a $I_\mathrm{RF}^2$ dependence, the 3D calculations provide also a means to estimate their effective resistance. From a scan without plasma (black diamonds in figure \ref{fig:LIBERTY34_1}b), we obtain $R_\mathrm{F} = 0.87$ $\Omega$ for the Faraday shield (lateral wall plus back disk) and $R_\mathrm{s} = 0.12$ $\Omega$ for the steel ring.  Adding the effective resistances of all the passive metallic parts we obtain a driver resistance in vacuum,
$$
 R^\mathrm{vac}_\mathrm{d} \equiv R_\mathrm{RF} + R_\mathrm{EM} + R_\mathrm{F}+ R_\mathrm{s} = 0.12 + 0.04 + (0.84+0.03) +0.12 = 1.15  ~\Omega.
$$
As shown in figure \ref{fig:LIBERTY34_1}b, the dissipation is slightly smaller in presence of the plasma (note the logarithmic scale). If we do the same power-law fits for the scans in figure \ref{fig:LIBERTY34_1}, we obtain
$$
 R_\mathrm{d} \equiv R_\mathrm{RF} + R_\mathrm{EM} + R_\mathrm{F}+ R_\mathrm{s} = 0.12 + 0.04 + (0.71+0.03) + 0.09 = 0.99 ~\Omega,
$$
a value that must be compared with experimental measurements \cite{jain2022investigation-o}, which yield $R_\mathrm{d}^\mathrm{exp} = 0.95$ $\Omega$. The estimates of $R_\mathrm{d}$ are important to set a feedback mechanism in transport calculations, where the RF heating is imposed through $I_\mathrm{RF}$, much like in the experiments. A fixed nominal power in the RF source must regulate the current in the RF coil according to the power share in the driver, including the plasma.

\section{Conclusions}

Previous 2D electromagnetic calculations for the ICP of the SPIDER device \cite{lopezbruna20212Delectromagnet} have been extended to 3D. The new calculation tool permits studying non axi-symmetric features of the ICP. In these first calculations we have explored the structure of the Faraday shield lateral wall, an important piece in the drivers of powerful ICP RF sources like those of SPIDER. Other conditions fixed, the dissipation in the Faraday shield cylindrical wall is stronger the smaller the clearance of its slits and the larger the number of slits for a given clearance. Since it is made of copper, its temperature affects also the dissipation due to the dependence of the electrical conductivity: the hotter the copper the more it dissipates. This is important because it establishes a positive feedback in possible hot spots in the Faraday shield structure. The dissipation will increase where the copper heats up, thus implying more local dissipation and still higher temperature. The present 3D calculations are restricted to a domain inside the driver that does not include the RF coil and electromagnetic shield. When combined with knowledge of the effective resistance in these two parts, the calculations allow for estimates of the PTE and the Faraday shield effective resistance. According to the calculations, the effective driver resistance during operation is $R_\mathrm{d}\approx 1$ $\Omega$ in good agreement with measured values and previous studies in SPIDER. Due to the large ohmic power losses in the Faraday shield, the PTE is found in the range $30$--$45$\% depending on the type of plasma. These results are also in agreement with studies in similar devices \cite{briefi2022diagnostics-of-}.

\section*{Acknowledgement}

We sincerely appreciate the help of J. M. Reynolds-Barredo and P. L. Garc\'ia-M\"uller on the numerical implementation of the problem.

This work has been carried out within the framework of the EUROfusion Consortium, funded by the European Union via the Euratom Research and Training Programme (Grant Agreement No 101052200 --- EUROfusion). Views and opinions expressed are however those of the author(s) only and do not necessarily reflect those of the European Union or the European Commission. Neither the European Union nor the European Commission can be held responsible for them.

\bibliographystyle{IEEEtran} % iopart-num, amsalpha, IEEEtran
%\bibliography{/Users/Daniel/Documents/NBTF}
\bibliography{olla}

\end{document}